# Exploring Co, Fe, and Ni Reference Layers for Single-Pulse All-Optical Reversal in Ferromagnetic Spin Valves

Jun-Xiao Lin, Yann Le Guen, Julius Hohlfeld, Jon Gorchon, Grégory Malinowski, Stéphane Mangin, Daniel Lacour, Thomas Hauet and Michel Hehn*

*Université de Lorraine, CNRS, Institut Jean Lamour, F-54000 Nancy, France*

*Authors to whom correspondence should be addressed: michel.hehn@univ-lorraine.fr

**Abstract**

We investigate the magnetization reversal process induced by a single femtosecond laser pulse in ferromagnetic spin valves by systematically comparing reference layers composed of pure Co, Ni, and Fe. To circumvent the loss of perpendicular magnetic anisotropy associated with changes in reference layer material and thickness, we design spin valves with in-plane magnetizations. While antiparallel-to-parallel switching is observed for all three elements, parallel-to-antiparallel switching occurs only with a Co reference layer and is absent with Ni and Fe. This difference is attributed to the distinct ultrafast magnetization dynamics of the reference materials. Our results support the hypothesis that parallel-to-antiparallel switching requires a rapid remagnetization of the reference layer, which generates a substantial negative spin current polarized opposite to the free layer magnetization—an essential condition for triggering its reversal.

# I. INTRODUCTION

The ultrafast manipulation of magnetization using femtosecond laser pulses offers a promising pathway to push spintronic devices and magnetic switching into the sub-picosecond regime [Kim19]. By driving magnetic systems far from equilibrium, purely longitudinal magnetization dynamics can circumvent the precessional constraints inherent to conventional switching schemes [Bea96, Tud04, Vah09].

However, this approach faces a fundamental speed limitation due to *critical slowing down* (CSD). As the magnetization approaches zero during the ferromagnetic–paramagnetic phase transition, its recovery rate dramatically decreases [Kaz08, Atx10, Man12], resulting in a pronounced slowdown in remagnetization. CSD can be mitigated in spin-valve structures where the layer targeted for reversal experiences a spin current generated by the ultrafast demagnetization of an adjacent magnetic layer [Bat10, Bee20, Bee23, Cho14]. This non-local transfer of spin angular momentum, mediated by a metallic spacer, has been exploited to trigger ultrafast magnetization reversal [Mal08, Rud12, Choi15], achieving switching times as short as a few hundred femtoseconds through ultrafast spin cooling [Hee20, Iil18, Rem20, Iga20, Rem22]. While such effects were initially demonstrated in systems with perpendicular magnetic anisotropy (PMA), they have recently been extended to in-plane magnetized spin valves [Lin24].

In 2023, Igarashi *et al.* reported single-pulse magnetization reversal in Gd-free spin valves consisting of [Co/Pt]$_3$/Cu/[Co/Pt]$_2$ layers, revealing unexpected parallel-to-antiparallel (P-to-AP) switching. This behavior was tentatively attributed to a spin-transfer-torque (STT)-like scattering mechanism at the Co/Cu interface [Iga23]. Building on this, Singh *et al.* provided time-resolved evidence that the spin current arising from the ultrafast magnetization dynamics of the reference layer is the primary driver of such reversal [Sin25].

Here, we extend this investigation to in-plane magnetized spin valves comprising a CoFeB free layer and different 3d transition-metal reference layers: Co, Ni, and Fe. These materials exhibit well-documented yet distinct ultrafast magnetization dynamics [Bor21, Sch23] and contrasting spin-polarization properties—positive in Fe and negative in Co and Ni [Kob22]. We show that antiparallel-to-parallel (AP-to-P) switching occurs in all three systems. In contrast, deterministic P-to-AP switching is observed only with a Co reference layer. Time-resolved magneto-optical Kerr effect (TR-MOKE) measurements and spin current calculations reveal that successful P-to-AP reversal requires rapid remagnetization of the reference layer, producing a spin current polarized opposite to the free-layer magnetization. These results highlight that the remagnetization dynamics—not merely demagnetization—of the reference layer play a critical role in enabling P-to-AP all-optical switching in spin-valve systems.

## II. RESULTS

### A. Magnetic properties

Controlling the magnetic properties and ensuring proper decoupling of ferromagnetic (FM) layers in spin valves becomes particularly challenging when comparing different FM materials. To address this, we adopted an in-plane magnetized spin-valve architecture, which offers greater flexibility in the choice and thickness of magnetic layers. To maintain consistency across all samples, we implemented a common reference subsystem based on an artificial antiferromagnetic (AAF) structure, known to provide enhanced magnetic rigidity compared to a single FM layer [Ber97]. The simplest AAF consists of two magnetic layers antiferromagnetically coupled via a thin metallic spacer. In the low-field regime, the AAF behaves as a rigid magnetic entity, with an effective magnetic rigidity gain defined as $Q = \frac{(m_1+m_2)}{(m_1-m_2)}$, where m1 and m2 are the magnetic moments of the individual layers. Strong antiferromagnetic (AF) coupling was achieved via the Ruderman-Kittel-Kasuya-Yosida (RKKY) interaction, using ruthenium (Ru) spacers, which exhibit high coupling strength as reported in [Par90].

The multilayer stacks fabricated for this study followed the structure: Si/SiO$_2$(500)/Ta(3)/Cu(5)/Co(3)/Ru(0.72)/FM(y)/Cu(5)/CoFeB(t)/Pt(3), where *FM* denotes Co, Fe, or Ni, and the numbers in parentheses indicate the layer thickness in nanometers. The base stack—Ta(3)/Cu(5)/Co(3)/Ru(0.72)—was identical across all samples, with the Ru thickness optimized to maximize AF coupling. The Co$_{40}$Fe$_{40}$B$_{20}$ free layer was chosen based on our previous study [Lin24], as its Curie temperature remains constant over the examined thickness range. For FM = Co and Ni, the reference-layer thickness *y* was fixed at 1.8 nm and 5 nm, respectively, while the CoFeB thickness *t* was varied from 1.3 to 2.2 nm. In contrast, for FM = Fe, the CoFeB thickness was fixed at *t* = 1.95 nm, and the reference-layer thickness *y* was varied between 1.5 and 2.0 nm.

| Material | Saturation magnetization (kA m$^{-1}$) | Curie temperature ($T_C$) (K) |
|---|---|---|
| Co | 1420 | 1300 |
| Fe | 1700 | 1040 |
| Ni | 490 | 620 |
| Co$_{40}$Fe$_{40}$B$_{20}$ | 1030 | 680 |

**TABLE. 1. Material parameters for Ni, Fe, Co, and Co$_{40}$Fe$_{40}$B$_{20}$.** Saturation magnetization and Curie temperature values for Co, Fe, and Ni are taken from the literature. The corresponding values for Co$_{40}$Fe$_{40}$B$_{20}$ are taken from [Lin24].

Figures S1–S3 show representative in-plane hysteresis loops measured along the easy axis using a commercial vibrating sample magnetometer. The full loops between positive and negative saturation confirm the characteristic behavior of our AAF stack: at low fields, the Co(3)/Ru(0.72)/FM(y) system reverses as a single, antiferromagnetically coupled entity with high coercivity, while the gradual slope toward saturation reflects the progressive rotation of the FM layer toward the applied field direction (Figure S4). In our AAF structure, the magnetic moment of the FM layer is always smaller than that of the Co(3) layer. As a result, under low applied fields, the Co magnetization remains aligned with the field, while the FM layer is antiparallel.

As shown in panels (b) of Figures S1–S3, minor loops measured under field amplitudes below 20 mT exhibit a plateau corresponding to the parallel alignment of the CoFeB and FM layers, which switches to the antiparallel configuration when the applied field exceeds approximately 15 mT. In all cases, the magnetization state of the CoFeB and FM layers—whether parallel or antiparallel—is reliably controlled by the field history. From the shifts observed in these minor loops (panels (b) of Figures S1–S3), we estimate the interlayer coupling fields between CoFeB and the reference layers to be below 0.58 mT for Co, 0.6 mT for Fe, and 0.5 mT for Ni.

**B. All-optical switching experiments**

All-optical switching experiments were carried out using a linearly s-polarized pump laser (1030 nm wavelength, 250 fs pulse duration) incident normal to the film and an s-polarized probe beam (670 nm wavelength) directed at 45° following a longitudinal Kerr geometry. Both beams illuminated the CoFeB side of the stack (Figure S5(a)) [Lin24]. The pump beam had a spot diameter of approximately 110-160 μm, while the probe beam was larger and operated at a sufficiently low energy to avoid affecting the magnetization switching behavior. Representative Kerr hysteresis loops measured along the in-plane easy axis for spin valves incorporating Co, Fe, or Ni as the reference FM layer are shown in Figure S5(b). For static single-pulse measurements, the sample magnetization was initially saturated using an external in-plane magnetic field—exceeding the coercivity and aligned along the easy axis—to set either the parallel or antiparallel configuration between the CoFeB free layer and the FM in the reference layers. After turning off the field, the remnant magnetization remained fixed along the in-plane easy axis. The samples were then exposed to varying numbers of pump pulses to evaluate their switching behavior.

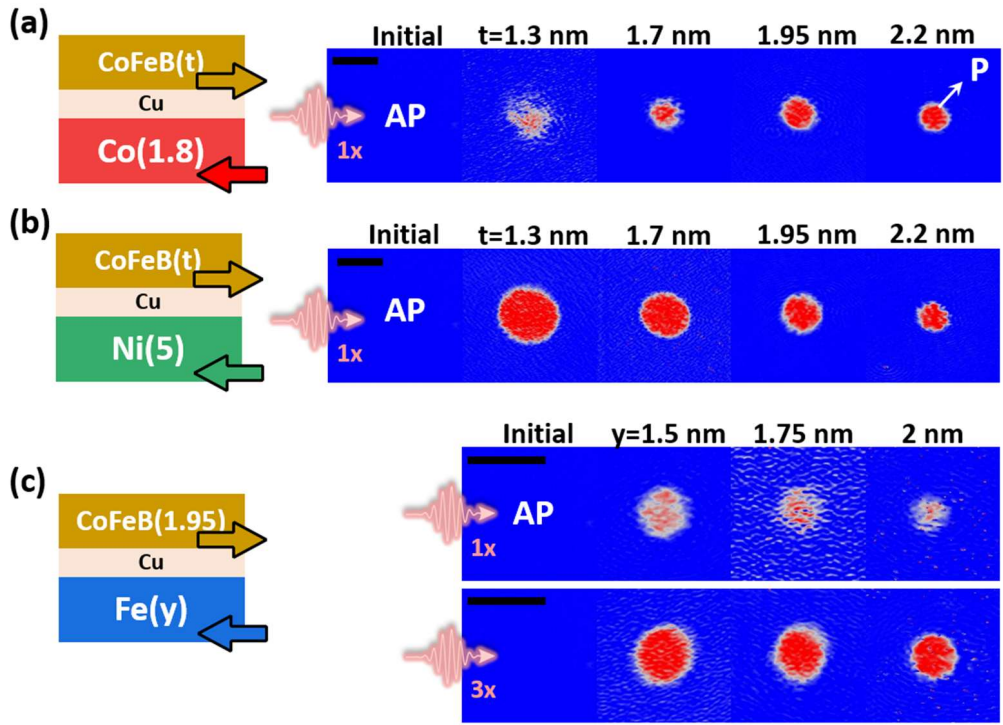

**FIG. 1. All-optical switching in in-plane magnetized spin valves, initialized in the antiparallel (AP) state.** The layer stacks are: $Si/SiO_2(500)/Ta(3)/Cu(5)/Co(3)/Ru(0.72)/FM(y)/Cu(5)/CoFeB(t)/Pt(3)$, where numbers in parentheses indicate thicknesses in nanometers. Kerr images were acquired after a single 250-fs laser pulse. (a) FM = Co(1.8 nm), with $t$ varied from 1.3 to 2.2 nm; laser fluence: 12.1 mJ·cm$^{-2}$. (b) FM = Ni(5 nm), with $t$ varied from 1.3 to 2.2 nm; laser fluence: 13.4 mJ·cm$^{-2}$. (c) FM = Fe, with $y$ varied from 1.5 to 2.0 nm and fixed $t$ = 1.95 nm; laser fluence: 16.1 mJ·cm$^{-2}$. The black scale bar in all Kerr images corresponds to 100 μm.

We first conducted experiments starting from the antiparallel (AP) configuration. Figure 1 compares the resulting magnetic states after a single laser pulse for each of the three FM reference layers. A single pulse induces a transition from the AP state (dark blue) to the P state

(dark red) for all cases except FM = Fe, which requires three successive pulses to achieve full reversal. Kerr images acquired after exposure to varying numbers of laser pulses are presented in Figure S6. This need for multiple pulses in Fe-based structures has been previously reported in in-plane magnetized systems [Lin24], and is attributed to a single spin current pulse delivering insufficient angular momentum to complete the switching.

In all FM cases studied, the reversal of the CoFeB layer can be attributed to a spin current pulse originating from the ultrafast demagnetization of the FM reference layer, with polarization opposite to the initial magnetization of the CoFeB free layer. This interpretation is consistent with earlier studies of Gd-based spin valves, where the Gd layer acts as the source of the spin current pulse [Xu17, Iil18].

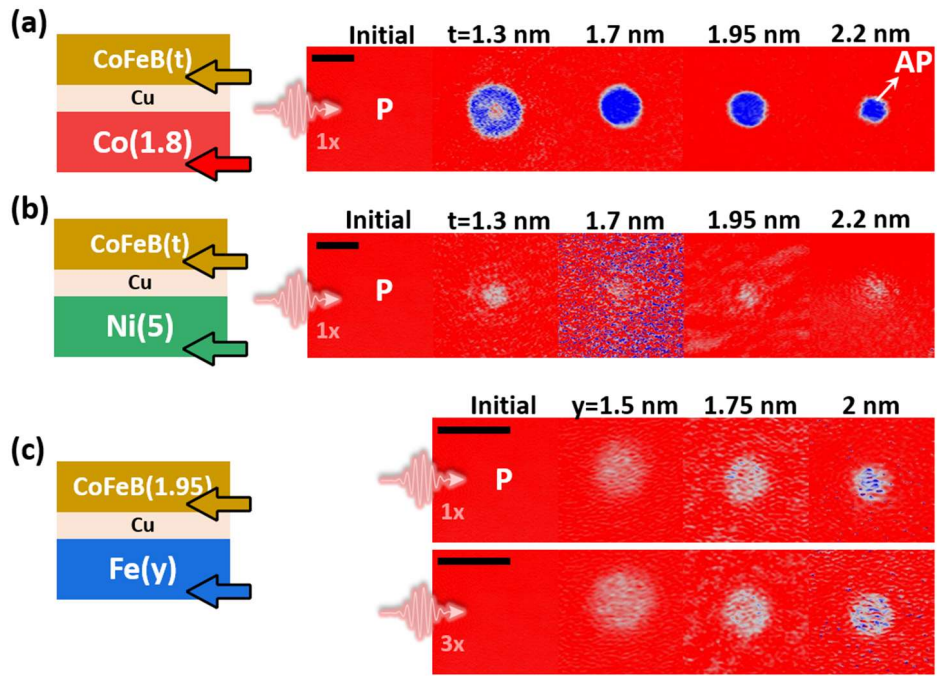

**FIG. 2. All-optical switching in in-plane magnetized spin valves, initialized in the parallel (P) state.** The layer stacks are: Si/SiO$_2$(500)/Ta(3)/Cu(5)/Co(3)/Ru(0.72)/FM(y)/Cu(5)/CoFeB(t)/Pt(3), where numbers in parentheses indicate thicknesses in nanometers. Kerr images were acquired after a single 250-fs laser pulse. (a) FM = Co(1.8 nm), with $t$ varied from 1.3 to 2.2 nm; laser fluence: 12.1 mJ·cm$^{-2}$. (b) FM = Ni(5 nm), with $t$ varied from 1.3 to 2.2 nm; laser fluence: 20.6 mJ·cm$^{-2}$. (c) FM = Fe, with $y$ varied from 1.5 to 2.0 nm and fixed $t$ = 1.95 nm; laser fluence: 22.9 mJ·cm$^{-2}$. The black scale bar in all Kerr images corresponds to 100 μm.

We next investigated magnetization switching from the parallel (P) state, where the magnetizations of the FM and CoFeB layers are initially aligned. Samples were prepared in the P configuration and then exposed to single 250-fs laser pulses. Figure 2 shows the resulting

Kerr images for samples with Co, Ni, and Fe as the reference layers. The final magnetic domain state depends strongly on the nature of the FM reference layer. For FM = Co, a single laser pulse induces a clear transition from the P state (dark red) to the AP state (dark blue), indicating complete switching of the CoFeB layer. Notably, this result reproduces the observation reported by Igarashi *et al.* in perpendicularly magnetized [Co/Pt]$_3$/Cu/[Co/Pt]$_2$ spin valves [Iga23].

In contrast, for FM = Ni and FM = Fe, even after up to five successive pulses, only a demagnetization of the CoFeB layer is observed, without any clear reversal (Figures S7-S8). We therefore conclude that P-to-AP reversal of the CoFeB layer occurs only when Co is used as the reference layer, and fails when Fe or Ni serves in that role.

To summarize the switching behavior, Figure 3 presents the experimental threshold fluences required to achieve magnetization reversal ($F_{th}^{Sw}$) and to reach a demagnetized state ($F_{th}^{Dem}$) as a function of reference-layer material, CoFeB thickness, and initial magnetization state. For antiparallel-to-parallel (AP-to-P) switching, the threshold fluence $F_{th\,(AP\,to\,P)}^{Sw}$ gradually increases from approximately 7.5 to 10 mJ·cm$^{-2}$ as the CoFeB thickness increases from 1.3 to 2.21 nm in both Co- and Ni-based stacks. For Fe, the threshold fluence required to reverse the 1.95 nm CoFeB layer is comparable to that of Co and Ni. These results are consistent with earlier findings [Lin24], which showed that successful switching requires sufficient reduction of the CoFeB magnetization at the time the spin current pulse arrives. Because the Curie temperature ($T_C$) of CoFeB remains constant across the studied thickness range (1.2–2.2 nm), increasing thickness requires proportionally higher fluences to reach the demagnetization threshold. Importantly, the AP-to-P threshold is largely independent of the FM material used as the reference layer, suggesting that the spin current generated during demagnetization is sufficient in all three cases to induce reversal when the CoFeB layer is adequately demagnetized.

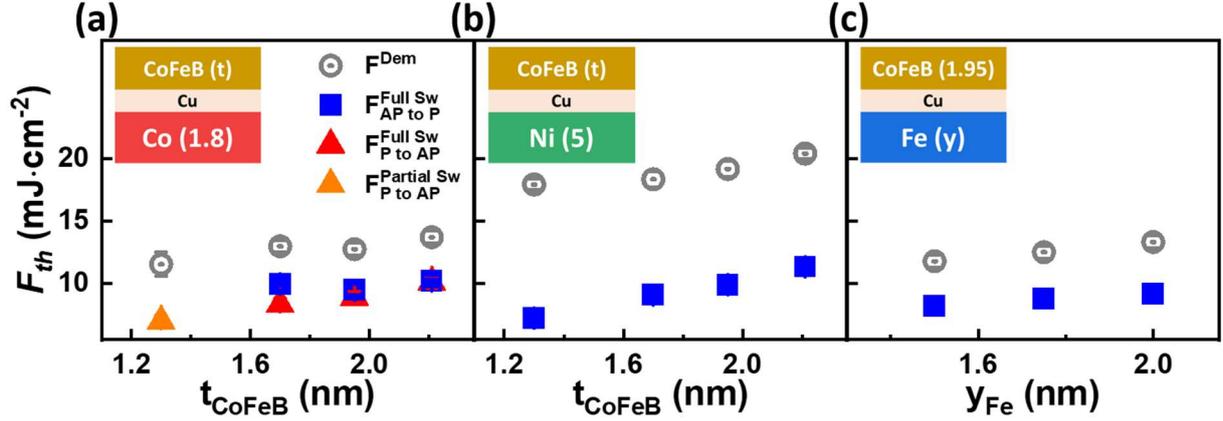

**FIG. 3. Experimental threshold fluences for magnetization reversal ($F_{th}^{Sw}$) and for transition to a demagnetized state ($F_{th}^{Dem}$).** Blue filled squares correspond to the AP-to-P transition; orange and red filled triangles represent the P-to-AP transition; and grey dotted circles indicate the threshold for reaching a demagnetized state. Fluences are plotted as a function of: (a) CoFeB thickness for a Co(1.8 nm) reference layer, (b) CoFeB thickness for a Ni(5 nm) reference layer, and (c) FM reference-layer material (Fe), at a fixed CoFeB thickness of 1.95 nm.

The threshold fluence for reaching a demagnetized state, $F_{th}^{Dem}$, exhibits a similar thickness dependence. It represents the condition under which the CoFeB layer remains in a demagnetized state either due to insufficient spin current or concurrent demagnetization of the FM reference layer. The $F_{th}^{Dem}$ values for Co and Fe are nearly identical, while those for Ni are approximately 1.5 times higher. This increased threshold in the Ni case can be attributed to the greater thickness of the Ni layer temperate by its lower $T_C$, which results in a larger thermal load required to induce sufficient demagnetization.

For Co-based spin valves, we also identify a threshold fluence for P-to-AP switching, $F_{th\ (P\ to\ AP)}^{Sw}$. As in the AP-to-P case, this fluence must be sufficient to significantly demagnetize the CoFeB layer, and therefore increases with its thickness. However, the precise threshold also depends on the remagnetization dynamics of the reference layer, which in turn are governed by its $T_C$. While one might expect identical thresholds for fixed CoFeB thickness, variations in ultrafast heating and cooling behavior across different FM layers result in distinct switching efficiencies.

As shown in Table 1, the $T_C$ of CoFeB (680 K) lies between that of Ni (620 K) and Fe (1040 K), with Co exhibiting the highest $T_C$ (1300 K). Consequently, the heating dynamics of the CoFeB layer and the resulting spin current contributions from the reference layer vary with the FM material, which ultimately influences the switching behavior.

## C. Time-resolved magnetization dynamics and calculated spin current

To compare the ultrafast magnetization dynamics of the different FM layers, we refer to trilayers of the form Glass/Ta(2 nm)/FM(12 nm)/Pt(2 nm), where the numbers in parentheses indicate layer thicknesses in nanometers. These structures were fabricated by magnetron sputtering and previously studied in [Sch23]. Time-resolved magneto-optical Kerr effect (TR-MOKE) measurements revealed a characteristic two-step magnetization response in all cases: an initial rapid drop in magnetization (demagnetization), followed by a slower recovery toward the equilibrium state (remagnetization). Figure 4 presents normalized TR-MOKE data recorded at a fixed laser fluence (panel a), and at fluences adjusted to yield the same demagnetization amplitude (panel b). Within the framework of the dM/dt model, the generation and polarity of spin current are directly proportional to the time derivative of the magnetization.

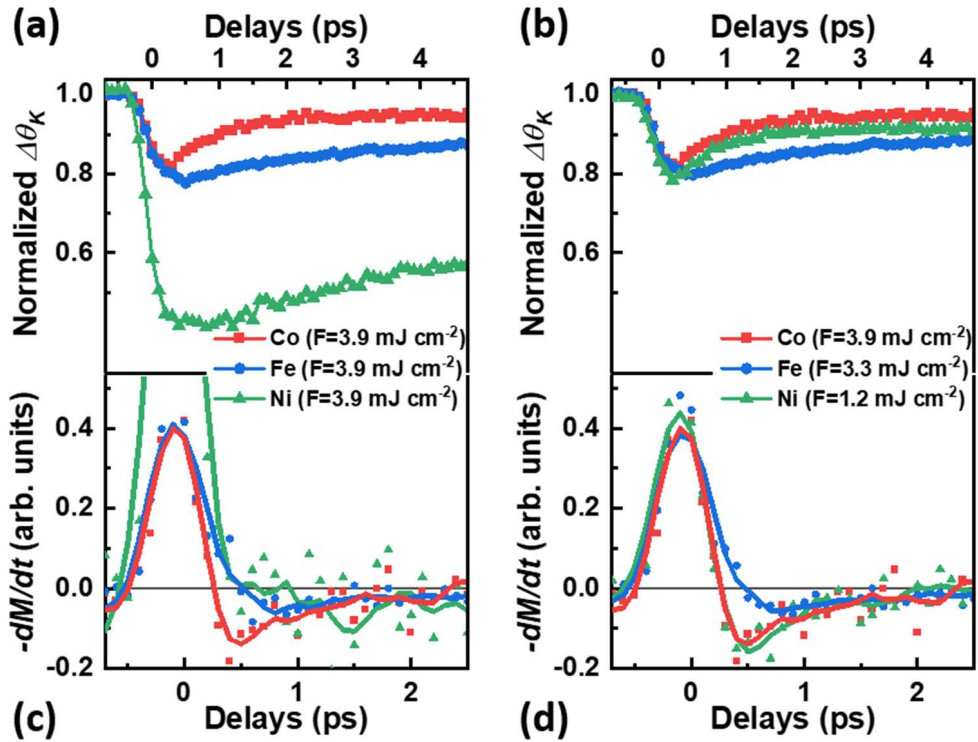

**FIG. 4. Time-resolved magneto-optical Kerr effect (TR-MOKE) measurements on Glass/Ta(2 nm)/FM(12 nm)/Pt(2 nm) trilayers, with FM = Co, Fe, or Ni.** Data are adapted from [Sch23]. (a) TR-MOKE traces recorded at a fixed pump fluence F = 3.9 mJ/cm$^2$. (b) TR-MOKE traces recorded at varied fluences, adjusted to produce the same demagnetization amplitude. (c, d) Corresponding spin current pulses, estimated from the negative time derivative of the magnetization (−dM/dt) of the data shown in (a) and (b), respectively. Filled symbols represent raw −dM/dt data; solid lines are Savitzky–Golay filtered fits (8-point window).

Therefore, to estimate the spin current $J_S$ emitted from each FM layer, we calculated -dM/dt from the TR-MOKE data. The resulting spin current pulse profiles are shown in Figures 4(c) and 4(d), corresponding to the experimental conditions in panels (a) and (b), respectively. Raw data are shown as filled symbols, while solid lines represent Savitzky–Golay filtered fits (8-point window).

At the same pump fluence, Ni exhibits the largest demagnetization, consistent with its lowest $T_C$ among the three materials, while Co shows the smallest demagnetization, reflecting its highest $T_C$ (Figure 4a). Both Fe and Ni display slower remagnetization dynamics compared to Co, resulting in predominantly unipolar spin current pulses. In contrast, Co generates a bipolar spin current profile—characterized by an initial positive peak followed by a longer-lasting negative tail (Figure 4c). As we will discuss later, this negative spin current plays a crucial role in enabling P-to-AP reversal.

When the fluence is adjusted such that each FM layer reaches the same demagnetization level (20%), Fe again shows the slowest recovery, yielding a strongly unipolar spin current pulse. In comparison, both Co and Ni produce bipolar profiles under these conditions as well (Figure 4d).

## III. DISCUSSION AND CONCLUSION

In this study, we demonstrated laser-induced magnetization switching in in-plane magnetized ferromagnetic (FM) spin-valve structures free of rare-earth elements, thereby extending the Gd-free all-optical switching (AOS) observations of Igarashi *et al.* [Iga23] to in-plane configurations. This geometry relaxes the constraints associated with perpendicular magnetic anisotropy and enables systematic investigation of both antiparallel-to-parallel (AP-to-P) and parallel-to-antiparallel (P-to-AP) reversal mechanisms across different FM reference materials.

We begin by examining the switching process from the AP to the P magnetic configuration. For a fixed CoFeB free-layer thickness, we find that the fluence threshold for AP-to-P reversal is largely independent of the FM material used in the reference layer. However, the degree of demagnetization of the reference layer varies significantly—being strongest for Ni, followed by Fe and Co—reflecting their respective Curie temperatures. Since the spin current driving the reversal is proportional to the negative time derivative of the magnetization, this implies that

Ni generates the largest spin-current pulse, thereby facilitating more efficient switching of the CoFeB layer. Additionally, the area of the reversed CoFeB domain systematically decreases with increasing free-layer thickness, confirming that thicker layers require more spin angular momentum to achieve full switching.

In contrast, P-to-AP switching is only observed when Co is used as the reference layer. This process requires a spin current polarized opposite to the initial magnetization of the CoFeB layer, which is only achieved if the reference layer exhibits a sufficiently rapid remagnetization after demagnetization. Time-resolved MOKE and spin current calculations (Figure 4) confirm that only Co produces a pronounced negative spin-current tail, essential for P-to-AP reversal. Ni and Fe, by contrast, display slower remagnetization dynamics and generate predominantly unipolar spin currents, which are insufficient to reverse the CoFeB layer from the parallel state. Moreover, the absence of divergent behavior between Fe (positive spin polarization at the Fermi level) and Co/Ni (negative) argues against a dominant role of STT-like scattering at the Cu/FM interface.

Our results suggest that by engineering the $T_C$ (e.g., via boron content) of the CoFeB free layer—or any soft layer targeted for switching—one could lower the threshold fluence required to reach its demagnetized state. This may enable P-to-AP switching with both Co and Ni reference layers, provided their remagnetization dynamics are sufficiently fast. Indeed, when the FM layers are driven to 20% demagnetization (Figure 4d), both Co and Ni exhibit negative spin-current contributions, indicating that such reversal could be achieved through careful thermal and structural tuning.

More generally, these findings reinforce the idea that achieving deterministic single-pulse switching—both AP-to-P and P-to-AP—requires matching the ultrafast spin dynamics of the reference and free layers [Zha25]. This interplay depends critically on their respective $T_C$ and magnetization dynamics. As shown here and in prior work [Iga23, Sin25], rare-earth-free all-optical switching can be realized in both in-plane and out-of-plane geometries using Co-based reference layers. From a broader technological perspective, this work highlights the relevance of $T_C$ and anisotropy energy (KV) scaling for AOS devices. While the threshold fluence for switching is closely tied to $T_C$, the stability of the final magnetic state scales with KV. This decoupling opens a promising pathway toward ultrafast and energy-efficient magnetic memories, where thermal switching thresholds can be reduced without sacrificing retention.

The results thus provide a foundational step toward scalable AOS-based spintronic devices with sub-picosecond performance and low power consumption.

**ACKNOWLEDGMENTS**

This work is supported by the ANR-20-CE09-0013 UFO, ANR-20-CE24-0003 SPOTZ, and ANR SLAM, the Institute Carnot ICEEL, the Région Grand Est, the Metropole Grand Nancy for the project "OPTIMAG" and FASTNESS, the interdisciplinary project LUE "MAT-PULSE", part of the French PIA project "Lorraine Université d'Excellence" reference ANR-15-IDEX-04-LUE, a European Union Program, the European Union's Horizon 2020 research and innovation program COMRAD under the Marie Skłodowska-Curie grant agreement No 861300. This article is based upon work from COST Action CA17123 MAGNETOFON and CA23136 CHIROMAG, supported by COST (European Cooperation in Science and Technology). This work was supported by the ANR through the France 2030 government grants EMCOM (ANR-22-PEEL-0009), PEPR SPIN (ANR-22-EXSP 0002) and PEPR SPIN – SPINMAT ANR-22-EXSP-0007. All funding was shared equally among all authors.

# Exploring Co, Fe, and Ni Reference Layers for Single-Pulse All-Optical Reversal in Ferromagnetic Spin Valves

Jun-Xiao Lin, Yann Le Guen, Julius Hohlfeld, Jon Gorchon, Grégory Malinowski, Stéphane Mangin, Daniel Lacour, Thomas Hauet and Michel Hehn*

*Université de Lorraine, CNRS, Institut Jean Lamour, F-54000 Nancy, France*

*Authors to whom correspondence should be addressed: michel.hehn@univ-lorraine.fr

## S1. Hysteresis loops for spin-valve structures employing Co as the reference layer

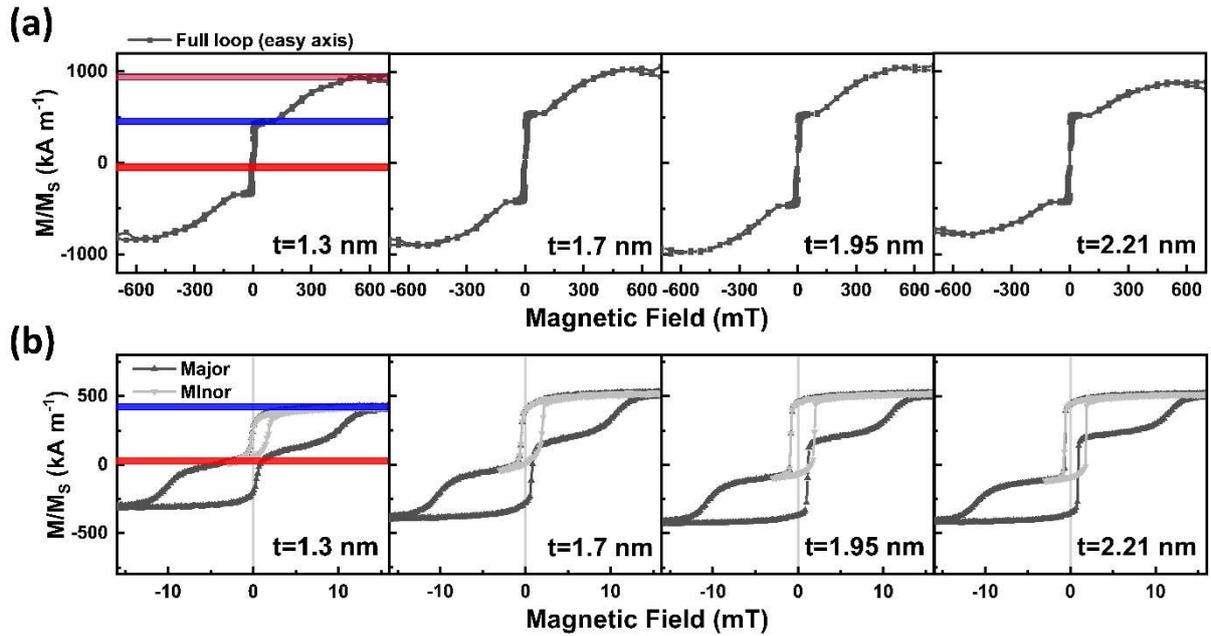

**FIG S1.** In-plane magnetic hysteresis loops measured for Si/ SiO(500)/Ta (3)/Cu (5)/Co (3)/Ru (0.72)/Co (1.8)/Cu (5)/CoFeB (t)/Pt (3). (a) Full-field loops recorded over a field range of ±800 mT. (b) Enlarged view of the low-field region (±20 mT), extracted from panel (a). The thick colored lines denote different relative magnetization configurations within the Co (3)/Ru (0.72)/Co (1.8)/Cu (5)/CoFeB (t) multilayers, as schematically illustrated in FIG S4.

## S2. Hysteresis loops for spin-valve structures employing Fe as the reference layer

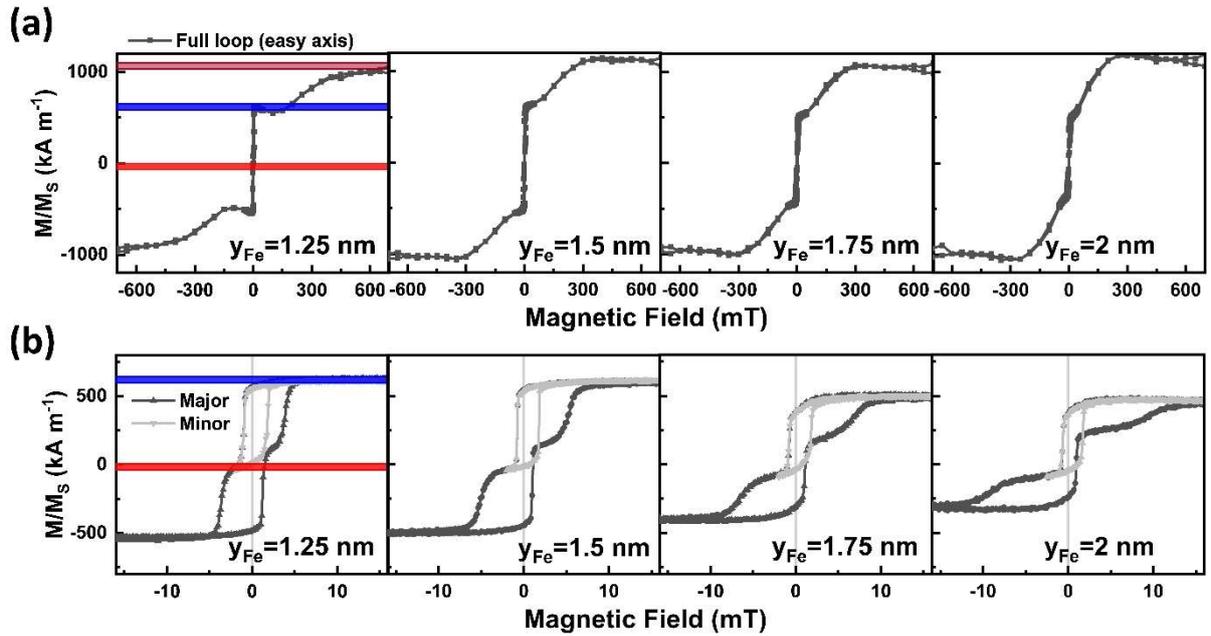

**FIG S2.** In-plane magnetic hysteresis loops for Si/ SiO(500)/Ta (3)/Cu (5)/Co (3)/Ru (0.72)/Fe (y)/Cu (5)/CoFeB (1.95)/Pt (3). (a) Hysteresis loops measured over a broad field range of ±800 mT. (b) Expanded view of the low-field region (±20 mT), extracted from panel (a). The thick colored lines indicate distinct relative magnetization configurations between the Co (3)/Ru (0.72)/Fe (y)/Cu (5)/CoFeB (1.95) layers.

## S3. Hysteresis loops for spin-valve structures employing Ni as the reference layer

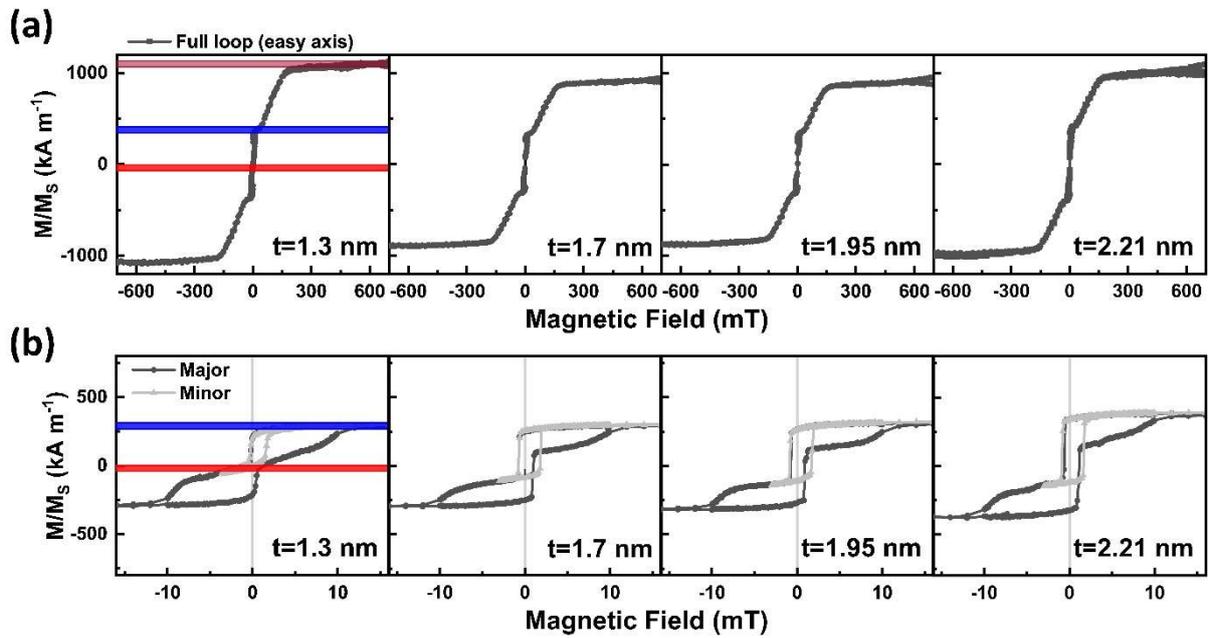

**FIG S3.** In-plane magnetic hysteresis loops for Si/ SiO(500)/Ta (3)/Cu (5)/Co (3)/Ru (0.72)/Ni (5)/Cu (5)/CoFeB (t)/Pt (3). (a) Full-field loops recorded between ±800 mT. (b) Zoomed-in view of the low-field region (±20 mT), extracted from the loops shown in (a). The thick colored lines correspond to distinct magnetization configurations among the Co (3)/Ru (0.72)/Ni (5)/Cu (5)/CoFeB (t) layers.

**S4. An example of moment configurations observed at different applied field strengths**

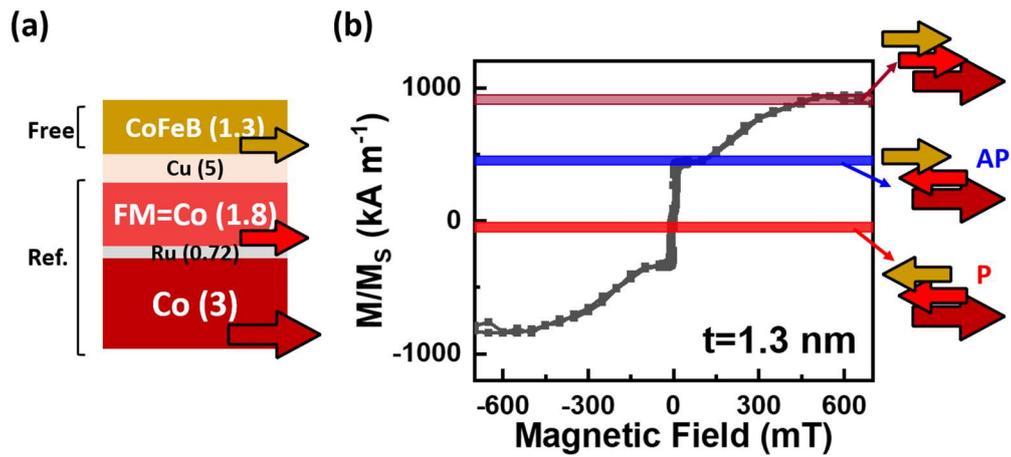

**FIG S4.** (a) Schematic representation of the multilayer stack, with arrows indicating the magnetization direction of each magnetic layer in the spin valve. (b) The definition of the antiparallel (AP) and parallel (P) magnetic states, which are consistently applied throughout the study for all spin valve samples with varying reference layer materials. This convention is valid because the magnetic moment of the FM layer is always smaller than that of the thicker Co (3 nm) layer in the reference stack.

## S5. Schematic of the experimental setup and the probed Kerr hysteresis loops

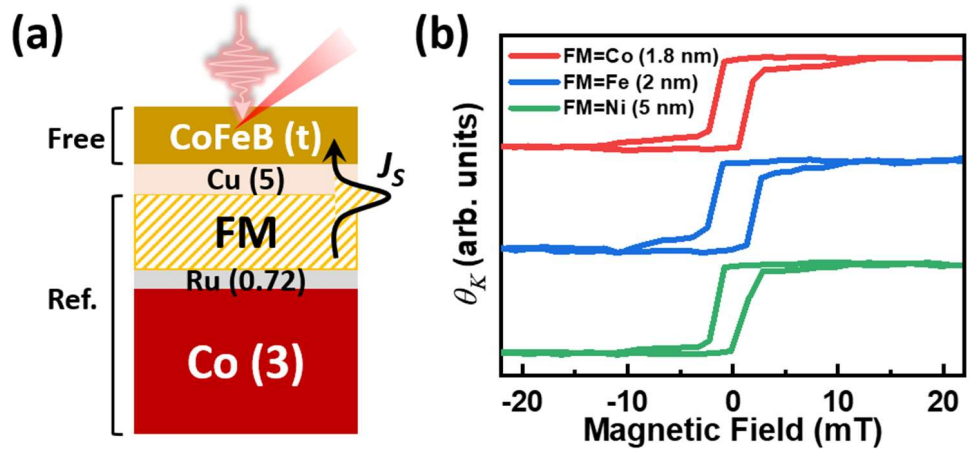

**FIG S5.** (a) Schematic of the sample structure and the static pump–probe experimental geometry. The pump beam is incident normal to the sample surface, while the probe beam approaches from the same side (i.e. the free-layer side) at an angle of 45°, enabling the acquisition of Kerr hysteresis loops and Kerr images following laser pulse excitation. The purpose of this study is also shown: to generate an optically induced spin current pulse ($J_S$) from the FM reference layer to switch the magnetization of the CoFeB free layer. (b) Representative Kerr hysteresis loops obtained under the configuration shown in (a) for spin valves incorporating Co, Fe, or Ni as the FM reference layer. The clear P and AP states are observed, following the definitions established in FIG. S4.

## S6. Magnetization manipulation under varying pulse numbers for spin valves with Fe reference layers, starting from the AP state

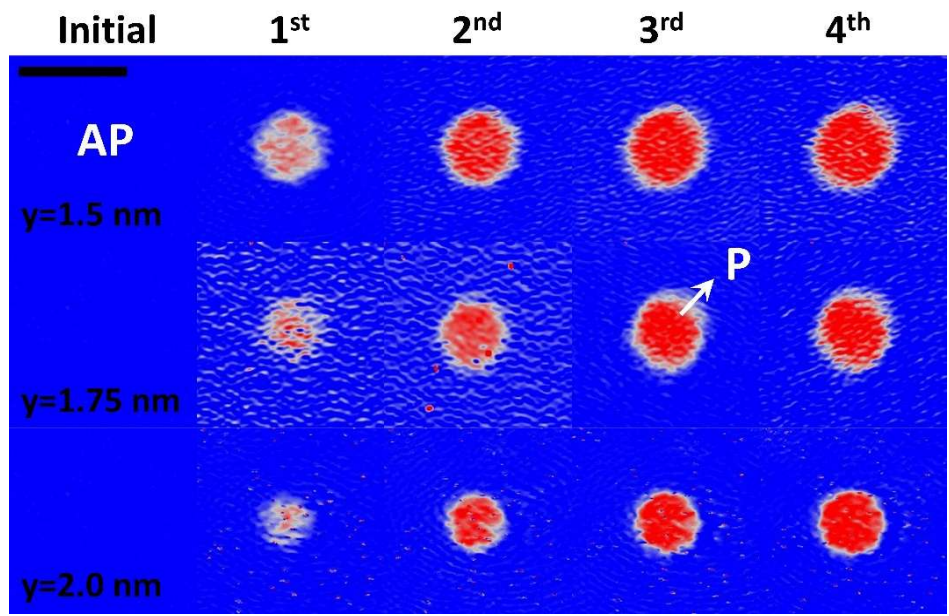

**FIG S6.** All-optical switching in in-plane magnetized spin valves, starting from the AP state. The layer stacks are Si/ SiO(500)/Ta (3)/Cu (5)/Co (3)/Ru (0.72)/Fe (y)/Cu (5)/CoFeB (1.95)/Pt (3), where y varied from 1.5 to 2 nm; laser fluence of 16.1 mJ cm$^{-2}$ is used. The scale bar in the Kerr images is 100 μm long.

## S7. Magnetization manipulation under varying pulse numbers for spin valves with Ni reference layers, starting from the P state

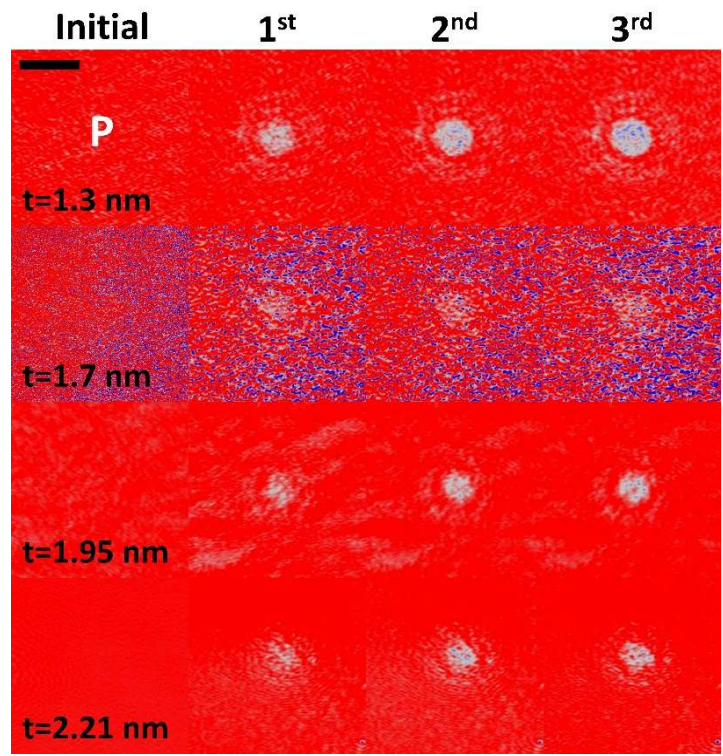

**FIG S7.** All-optical switching in in-plane magnetized spin valves, starting from the P state. The layer stacks are Si/ SiO(500)/Ta (3)/Cu (5)/Co (3)/Ru (0.72)/Ni (5)/Cu (5)/CoFeB (t)/Pt (3), where t varied from 1.3 to 2.21 nm; laser fluence of 20.6 mJ cm$^{-2}$ is used. The scale bar in the Kerr images is 100 μm long.

## S8. Magnetization manipulation under varying pulse numbers for spin valves with Fe reference layers, starting from the P state

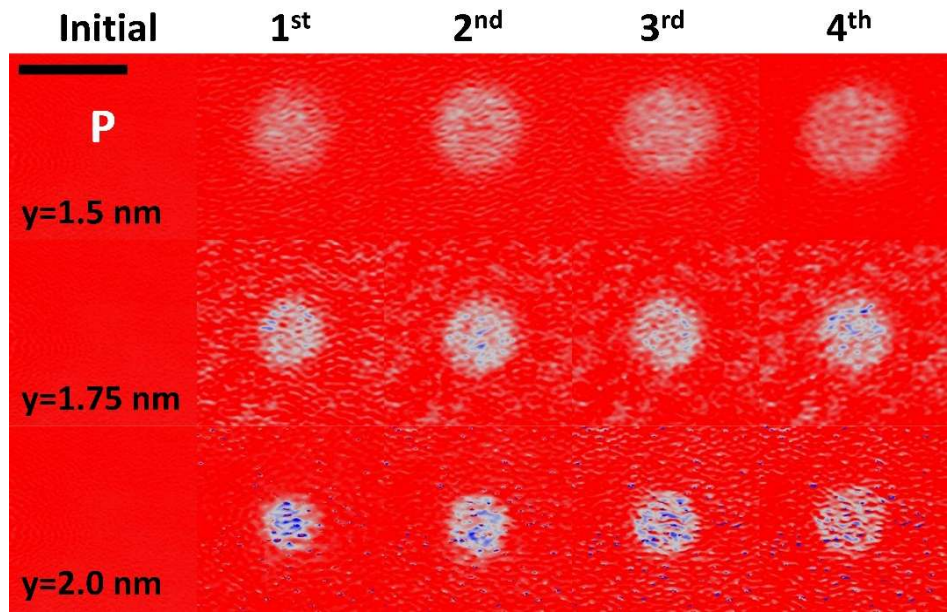

**FIG S8.** All-optical switching in in-plane magnetized spin valves, starting from the P state. The layer stacks are Si/ SiO(500)/Ta (3)/Cu (5)/Co (3)/Ru (0.72)/Fe (y)/Cu (5)/CoFeB (1.95)/Pt (3), where y varied from 1.5 to 2 nm; laser fluence of 22.9 mJ cm$^{-2}$ is used. The scale bar in the Kerr images is 100 μm long.